\documentclass[epj]{svjour}
\usepackage{graphicx}
\usepackage{amsmath}
\begin{document}
\title{Extended Hubbard model in the presence of a magnetic field}
\author{F. Mancini\inst{1} \and F. P. Mancini\inst{2}
} \institute{Dipartimento di Fisica ``{\it E. R. Caianiello}'' -
Unit\`a CNISM di Salerno, Universit\`a degli Studi di Salerno, Via
S. Allende, I-84081 Baronissi (SA), Italy. \and Dipartimento di
Fisica and Sezione I.N.F.N., Universit\`a degli Studi di Perugia,
Via A. Pascoli, I-06123 Perugia, Italy.}
%
%
\abstract{ Within the Green's function and equations of motion
formalism it is possible to exactly solve a large class of models
useful for the study of strongly correlated systems. Here, we
present the exact solution of the one-dimensional extended Hubbard
model with on-site $U$ and first nearest neighbor repulsive $V$
interactions in the presence of an external magnetic field $h$, in
the narrow band limit. At zero temperature our results establish
the existence of four phases in the three-dimensional space ($U,
n, h$) - $n$ is the filling - with relative phase transitions, as
well as different types of charge ordering. The magnetic field may
dramatically affect the behavior of thermodynamic quantities,
inducing, for instance, magnetization plateaus in the
magnetization curves, and a change from a single to a double-peak
structure in the specific heat. According to the value of the
particle density, we find one or two critical fields, marking the
beginning of full or partial polarization. A detailed study of
several thermodynamic quantities is also presented at finite
temperature.
} 

\maketitle

\section{Introduction}

In physics exact solutions are of great importance since, in some
cases, an approximation may introduce dramatically dominating
errors, resulting in an incorrect description of the phenomenon
under study. Recently, within the Green's function and equations
of motion formalism, we have shown that it is possible to exactly
solve a large class of models useful for the study of strongly
correlated systems \cite{Mancini05_06,mancini2}. By exactly
solvable we mean that it is always possible to find a set of
eigenenergies and eigenoperators of the Hamiltonian closing the
hierarchy of the equations of motion. Thus, one can obtain exact
expressions for the relevant Green's and correlation functions in
terms of a finite set of parameters \cite{manciniavella}.

The aim of the present paper is twofold. First, we would like to
further develop our previous work on the exact solution of the
one-dimensional extended Hubbard model in the atomic limit
(AL-EHM)  \cite{mancini2}, by extending it to a more general
situation in which a finite magnetization may be induced by an
external magnetic field. Secondly, the AL-EHM exhibits interesting
features at low temperatures, and one of the most fascinating
characteristic feature is that it shows magnetic plateaus.
Interestingly, magnetization plateaus have been predicted also for
spin-one bosons in optical lattices \cite{imambekov04}. By atomic
limit, according to the conventional definition used in the
literature, we mean the classical limit of the model, i.e., we set
from the very beginning the hopping matrix elements $t_{ij} = 0$.
One may obtain different results by treating the problem with
nonzero hopping and then taking the limit to zero.

A uniform magnetic field, introduced through a Zeeman term, has
dramatic effects both on the $T=0$ phase diagram and on the
behavior of several thermodynamic quantities. We study the
properties of the system as functions of the external parameters
$n$, $T/V$, $U/V$ and $h/V$ (throughout the paper we set $V=1$ as
the unit of energy), allowing for the on-site interaction $U$ to
be both repulsive and attractive. In fact, the parameter $U$ can
represent the effective interaction coupling taking into account
also other interactions.
Owing to the particle-hole symmetry, it is sufficient to explore
the interval $[0,1]$ for the parameter $n$. The chemical potential
is self-consistently determined as a function of the external
parameters. We address the problem of determining the
zero-temperature phase diagram in the ($U$, $n$, $h$) space and we
find various phase transitions (PTs), as well as magnetization
plateaus. At $T=0$, for attractive on-site interactions, the
magnetic field does not play any role if its intensity is $h<\vert
U \vert/2$: the ground state is a collection of doublons (sites
with two electrons of opposite spin), and there are no neighbor
sites occupied. The magnetic energy is not strong enough to break
the doublons. On the other hand, for strong repulsive on-site
interactions, it is sufficient a small nonzero value of the
magnetic field to have a finite magnetization. In the intermediate
regions, the competition among $U$, $V$ and $h$ determines the
phase structure.

Relevant thermodynamic quantities, such as the double occupancy,
the charge and spin susceptibilities, the specific heat and the
entropy are systematically computed both at $T=0$ and as functions
of the temperature. For all values of the particle density, one
finds a critical value (dramatically depending on $U$ and $n$) of
the magnetic field $h_s$ above which the ground state is
ferromagnetic. In this state, every occupied site contains one and
only one electron, aligned along the direction of $h$. For strong
repulsive on-site interactions, one finds $h_s=0$, i.e., the spin
are polarized as soon as the magnetic field is turned on.
Furthermore, for attractive on-site interactions and $0.5<n \le
1$, one observes the existence of two critical fields, namely:
$h_c$, up to which no magnetization is observed, and $h_s$,
marking the beginning of full polarization. This is analogous to
the finite field behavior of the $S=1$ Haldane chain
\cite{haldane}.

The addition of a homogeneous magnetic field does not dramatically
modify the framework of calculation given in Ref. \cite{mancini2},
provided one takes into account the breakdown of the spin
rotational invariance. For the sake of comprehensiveness, in
Section \ref{sec_II}, we briefly report the analysis leading to
the algebra closure and to analytical expressions of the retarded
Green's functions (GFs) and correlation functions (CFs). The GFs
and CFs depend on a set of internal parameters, which can be
determined by means of algebra constraints
\cite{Mancini05_06,manciniavella}, allowing us to provide an exact
and complete solution of the one-dimensional AL-EHM in the
presence of a magnetic field. In Section \ref{sec_III} we analyze
the properties of the system at zero temperature. We characterize
the different phases emerging in the phase diagram drawn in the
($U,n,h$) space, by studying the behavior of the chemical
potential and of various local properties (double occupancy,
short-range correlation function, magnetization); Section
\ref{sec_IV} is devoted to the study of the finite temperature
properties. Further to the study of the quantities analyzed in
Section \ref{sec_III}, we also present results for the charge and
spin susceptibilities, the specific heat and the entropy. Finally,
Sec. \ref{sec_V} is devoted to our conclusions and final remarks,
while the Appendix reports some relevant computational details.

\section{The model}
\label{sec_II}

The one-dimensional extended Hubbard model in the presence of an
external homogeneous magnetic field is described by the following
Hamiltonian
\begin{equation}
\label{eq1}
 \begin{split}
H&=\sum_{ij} [t_{ij} -\delta _{ij} \mu ]c^\dag (i)c(j)+U\sum_i
{n_\uparrow (i)n_\downarrow (i)}
\\
&+\frac{1}{2}\sum_{i\ne j} {V_{ij} n(i)n(j)} -h\sum_i n_3 (i).
\end{split}
\end{equation}
$n(i)=c^\dag (i)c(i)$ is the charge density operator, $c(i)$
$(c^\dag (i))$ is the electron annihilation (creation) operator -
in the spinor notation - satisfying canonical anti-commutation
relations. $n_3 (i)$ is the third component of the spin density
operator, also called the electronic Zeeman term,
\begin{equation}
\label{eq2} n_3 (i)=n_\uparrow (i)-n_\downarrow (i)=c_\uparrow
^\dag (i)c_\uparrow (i)-c_\downarrow ^\dag (i)c_\downarrow (i).
\end{equation}
Here we do not consider the orbital interaction with the magnetic
field and we use the Heisenberg picture: $i=({\bf i},t)$, where
{\bf i} stands for the lattice vector ${\bf R}_i$. The Bravais
lattice is a linear chain of $N$ sites with lattice constant $a$.
$\mu$ is the chemical potential; $h$ is the intensity of the
external magnetic field and $U$ and $V$ are the strengths of the
local and intersite interactions, respectively. In the atomic
limit, if one considers only first neighboring sites interactions,
the Hamiltonian \eqref{eq1} takes the form
\begin{equation}
\label{eq5}
 H= \sum_i \left[- \mu  n(i)+U  {D(i)} +\frac{1}{2} V
n(i)n^\alpha -h  n_3 (i)  \right],
\end{equation}
where $D(i)$ is the double occupancy operator, defined as $
D(i)=n_\uparrow (i)n_\downarrow (i)=n(i)[n(i)-1]/2$. Hereafter,
for a generic operator $\Phi(i)$, we define $\Phi ^\alpha
(i,t)=\sum_j \alpha _{ij} \Phi (j,t)$, where $\alpha_{ij}$ is the
projection operator over first nearest neighboring sites.

\subsection{Composite fields and equations of motion}

By taking time derivatives of increasing order of the fermionic
field $c(i)$, the dynamics generates other field operators of
higher complexity (composite operators). However, the number of
composite operators is finite because of the recurrence relation
\begin{equation}
\label{eq7}
 [n^\alpha (i)]^k=\sum_{m=1}^4 A_m^{(k)} [n^\alpha
(i)]^m ,
\end{equation}
which allows one to write the higher-power expressions of the
operator $n^\alpha (i)$ in terms of the first four powers. The
coefficients $A_m^{(k)}$  are rational numbers satisfying the
relations $\sum_{m=1}^4 A_m^{(k)}=1$ and $A_m^{(k)}=\delta_{m,k}$
($k=1,\ldots,4$), explicitly given in Ref. \cite{mancini05b}. As a
result, a complete set of eigenoperators of the Hamiltonian
\eqref{eq5} can be found.

Upon introducing the Hubbard operators $\xi(i)=[n(i)-1]c(i)$ and
$\eta (i)=n(i)c(i)$, one may define the composite field operator
\begin{equation}
\label{eq8} \psi (i)=
\begin{pmatrix}
\psi^{(\xi )}(i)  \\
\psi^{(\eta )}(i)
\end{pmatrix}
=
\begin{pmatrix}
\psi_{\uparrow} ^{(\xi )}(i)  \\
\psi_{\downarrow} ^{(\xi )}(i)  \\
\psi_{\uparrow} ^{(\eta )}(i)\\
\psi_{\downarrow} ^{(\eta )}(i)
\end{pmatrix},
\end{equation}
where
\begin{equation}
\label{eq10} \psi_\sigma ^{(\xi )}(i)=
\begin{pmatrix}
 \xi_\sigma (i)  \\
 \xi_\sigma (i)[n^\alpha (i)]  \\
 \vdots   \\
\xi_\sigma (i)[n^\alpha (i)]^4
\end{pmatrix},
\;
 \psi_\sigma^{(\eta )}(i)=
\begin{pmatrix}
 \eta_\sigma (i)  \\
 \eta_\sigma (i)[n^\alpha (i)]  \\
 \vdots   \\
\eta_\sigma (i)[n^\alpha (i)]^4
\end{pmatrix},
\end{equation}
where $\sigma=\{\uparrow,\downarrow \}$. With respect to the case
of zero magnetic field \cite{mancini2}, the degrees of freedom
have doubled, since one has to the take into account the two
nonequivalent directions of the spin. By exploiting the algebraic
properties of the operators $n(i)$ and $D(i)$, it is easy to show
that the fields $\psi ^{(\xi )}(i)$ and $\psi ^{(\eta )}(i)$ are
eigenoperators of the Hamiltonian \eqref{eq5} \cite{mancini05b}:
\begin{equation}
\label{eq14}
\begin{split}
 i\frac{\partial }{\partial t}\psi ^{(\xi )}(i)&=[\psi ^{(\xi
)}(i),H]=\varepsilon ^{(\xi )}\psi ^{(\xi )}(i) ,\\
 i\frac{\partial }{\partial t}\psi ^{(\eta )}(i)&=[\psi ^{(\eta
)}(i),H]=\varepsilon ^{(\eta )}\psi ^{(\eta )}(i).
 \end{split}
\end{equation}
$\varepsilon ^{(\xi )}$ and $\varepsilon ^{(\eta )}$ are the
$10\times 10$  energy matrices:
\begin{equation}
\label{eq15} \varepsilon ^{(\xi )}=\left( {{\begin{array}{*{20}c}
 {\varepsilon _\uparrow ^{(\xi )} }  & 0  \\
 0  & {\varepsilon _\downarrow ^{(\xi )} }  \\
\end{array} }} \right),
\qquad  \varepsilon ^{(\eta )}=\left(
{{\begin{array}{*{20}c}
 {\varepsilon _\uparrow ^{(\eta )} }  & 0  \\
 0  & {\varepsilon _\downarrow ^{(\eta )} }  \\
\end{array} }} \right),
\end{equation}
where
\begin{equation}
\label{eq16}
 \varepsilon _\sigma ^{(\xi )} =
\begin{pmatrix}
-\scriptstyle  \mu -\sigma h  & \scriptstyle 2V  &  \scriptstyle 0  & \scriptstyle 0  & \scriptstyle 0  \\
\scriptstyle 0  & \scriptstyle -\mu -\sigma h  & \scriptstyle 2V & \scriptstyle 0  &  \scriptstyle 0  \\
\scriptstyle 0  & \scriptstyle 0  & \scriptstyle -\mu -\sigma h  & \scriptstyle 2V  & \scriptstyle 0  \\
\scriptstyle 0  & \scriptstyle 0  & \scriptstyle  0  & \scriptstyle -\mu -\sigma h  & \scriptstyle 2V  \\
\scriptstyle 0  & \scriptstyle -3V  & \scriptstyle \frac{25}{2}V &
-\scriptstyle \frac{35}{2}V & \scriptstyle -\mu -\sigma h+10V
\end{pmatrix},
\end{equation}
\begin{equation}
\label{eq17} \varepsilon _\sigma ^{(\eta )} =
\begin{pmatrix}
 \scriptstyle  U -\mu -\sigma h  & \scriptstyle 2V  & \scriptstyle 0  & \scriptstyle 0  & \scriptstyle 0  \\
\scriptstyle 0  &  \scriptstyle U -\mu -\sigma h  & \scriptstyle 2V  & \scriptstyle 0  & \scriptstyle 0  \\
 \scriptstyle 0  & \scriptstyle 0  & \scriptstyle U -\mu -\sigma h  & \scriptstyle 2V  & \scriptstyle 0  \\
\scriptstyle 0  & \scriptstyle 0  & \scriptstyle 0  & \scriptstyle U -\mu -\sigma h  & \scriptstyle 2V  \\
\scriptstyle 0  & \scriptstyle -3V  & \scriptstyle \frac{25}{2}V &
\scriptstyle -\frac{35}{2}V & \scriptstyle U -\mu -\sigma h+10V
\end{pmatrix}.
\end{equation}
The eigenvalues of the matrices $\varepsilon ^{(\xi )}$ and
$\varepsilon ^{(\eta )}$ are
\begin{equation}
\label{eq18}
E^{(\xi )}=
\begin{pmatrix}
E_\uparrow ^{(\xi)}  \\
E_\downarrow ^{(\xi)}
\end{pmatrix},
 \qquad
 E^{(\eta )}=
\begin{pmatrix}
E_\uparrow ^{(\eta )}   \\
E_\downarrow ^{(\eta )}
\end{pmatrix},
\end{equation}
where
\begin{equation}
\label{EHM_7}
\begin{split}
 E^{(\xi )}_{p,\sigma}&=-\mu -\sigma h +(p-1)V,     \\
 E_{p,\sigma}^{(\eta )}&=U-\mu -\sigma h +(p-1)V ,
\end{split}
\end{equation}
with $p=1,\ldots ,5$. Although at the level of equations of motion
the two fields $\psi ^{(\xi )}(i)$ and $\psi ^{(\eta )}(i)$ are
decoupled, they are indeed coupled by means of the self-consistent
equations which determine the correlators appearing in the
normalization matrix, as shown in Subsection \ref{sse}.

\subsection{Retarded Green's and correlation functions}

The knowledge of a complete set of eigenoperators and
eigenenergies of the Hamiltonian \eqref{eq5} allows one to exactly
determine the retarded thermal Green's function
\begin{equation}
\label{eq20} G^{(a)}(t-t') =\theta (t-t')\langle\{\psi
^{(a)}(i,t), {\psi^{a}}^\dag (i,t')\}\rangle.
\end{equation}
In the above equations, $a=\xi ,\eta$ and $\langle\cdots \rangle$
denotes the quantum-statistical average over the grand canonical
ensemble. By means of the field equations (\ref{eq14}), it is easy
to show that the Green's function satisfies the equation
\begin{equation}
\label{eq21} [\omega -\varepsilon ^{(a)}]G^{(a)}(\omega )=I^{(a)},
\end{equation}
where $I^{(a)}$ is the normalization matrix
\begin{equation}
\label{eq22} I^{(a)}=\langle \{\psi ^{(a)}(i,t),{\psi ^{(a)}}^\dag
(i,t)\}\rangle,
\end{equation}
whose  expression in discussed in the Appendix. The solution of
Eq. \eqref{eq21} is
\begin{equation}
\label{eq27}
G^{(a)}(\omega )=\sum_{p=1}^{10} \frac{\sigma
^{(a,p)}}{\omega -E_p^{(a)} +i\delta },
\end{equation}
where the spectral functions $\sigma _{\mu \nu }^{(a,p)}$ can be
computed by means of the formula
\begin{equation}
\label{eq28} \sigma _{\mu \nu }^{(a,p)} =\Omega _{\mu p}^{(a)}
\sum_{\delta} {\Omega _{p\delta }^{(a)}}^{-1} I_{\delta \nu
}^{(a)}.
\end{equation}
$\Omega ^{(a)}$ is the $10\times 10$ matrix whose columns are the
eigenvectors of the matrix $\varepsilon ^{(a)}$. Explicit
expressions of the matrices $\Omega ^{(\xi )}=\Omega ^{(\eta )}$
are reported in the Appendix. The correlation function
\begin{equation}
\label{eq31} \begin{split} C^{(a)}(t-t')&=\langle \psi
^{(a)}(i,t){\psi ^{(a)}}^\dag (i,t')\rangle \\
&=\frac{1}{(2\pi )}\int_{-\infty }^{+\infty } d\omega e^{-i\omega
(t-t')}C^{(a)}(\omega )
\end{split}
\end{equation}
can be immediately computed from Eq. \eqref{eq27} by means of the
spectral theorem, and one finds
\begin{equation}
\label{eq32} C^{(a)}(\omega )=\pi \sum_{p=1}^{10} \left[ {1+\tanh
\left( {\frac{\beta \omega }{2}} \right)} \right]\sigma
^{(a,p)}\delta [\omega -E_p^{(a)} ],
\end{equation}
where $\beta =1/k_B T$. At equal time, the CF is given by
\begin{equation}
\label{eq33}
 C^{(a)}=\langle \psi ^{(a)}(i){\psi ^{(a)}}^\dag
(i)\rangle =\frac{1}{2}\sum_{p=1}^{10} T_p^{(a)} \sigma ^{(a,p)},
\end{equation}
where $ T_p^{(a)} =1+\tanh( E_p^{(a)}/2k_B T)$.  Upon introducing
the parameters
\begin{equation}
\label{eq35}
\begin{split}
 n&=\langle n(i)\rangle , \\
 m&=\langle n_3(i)\rangle , \\
 \kappa ^{(p)}&=\langle [n^\alpha (i)]^p\rangle,
 \end{split}
 \qquad
 \begin{split}
 \lambda ^{(p)}&=\frac{1}{2}\langle n(i)[n^\alpha (i)]^p\rangle  , \\
 \pi ^{(p)}&=\frac{1}{2}\langle n_3 (i)[n^\alpha
(i)]^p\rangle ,
\end{split}
\end{equation}
the elements of the normalization matrix can be written as
\begin{equation}
\label{eq36}
\begin{split}
I_{1,p}^{(\xi )} &=\kappa ^{(p-1)}-\lambda ^{(p-1)}+\pi ^{(p-1)} ,\\
I_{6,p+5}^{(\xi  )} &=\kappa ^{(p-1)}-\lambda ^{(p-1)}-
\pi^{(p-1)}, \\
I_{1,p}^{(\eta  )} &=\lambda ^{(p-1)}-\pi ^{(p-1)}, \\
I_{6,p+5}^{(\eta  )} &=\lambda ^{(p-1)}+ \pi^{(p-1)},
\end{split}
\end{equation}
where and $p=1,\ldots,5$. All the other matrix elements can be
easily computed by means of the recurrence relation \eqref{eq7}.

\subsection{Self-consistent equations}
\label{sse}

The previous analysis shows that the complete solution of the
model requires the knowledge of the following 13 parameters:
$\mu$, $m$, $\kappa ^{(2)},\ldots,\kappa ^{(4)}$, $\lambda
^{(1)},\ldots,\lambda ^{(4)}$, $\pi^{(1)},\ldots,\pi ^{(4)}$.
These quantities may be computed by using algebra constraints and
symmetry requirements \cite{Mancini05_06,manciniavella}. By
recalling the projection nature of the Hubbard operators $\xi(i)$
and $\eta(i)$, it is straightforward to verify that the following
algebraic properties hold
\begin{equation}
\label{eq38}
\begin{split}
\xi ^\dag (i)n(i)&=0 ,\\
\xi ^\dag (i)D(i)&=0,
\end{split}
\qquad
\begin{split}
\eta ^\dag (i)n(i)&=\eta ^\dag (i) ,\\
\eta ^\dag (i)D(i)&=0.
\end{split}
\end{equation}
These are fundamental relations and constitute the basis to
construct a self-consistent procedure to compute the various
parameters of the model. Upon splitting the Hamiltonian
(\ref{eq5}) as the sum of two terms:
\begin{equation}
\label{eq39}
\begin{split}
 H&=H_0 +H_I  ,\\
 H_I &=2Vn(i)n^\alpha (i),
 \end{split}
\end{equation}
the statistical average of any operator $O$ can be expressed as
\begin{equation}
\label{eq40} \langle O\rangle =\frac{\langle Oe^{-\beta
H_I}\rangle _0 }{\langle e^{-\beta H_I}\rangle _0 },
\end{equation}
where $\langle \cdots \rangle _0$ stands for the trace with
respect to the reduced Hamiltonian $H_0$: $\langle \cdots \rangle
_0 ={\rm Tr}\{\cdots e^{-\beta H_0}\}/  {\rm Tr}\{e^{-\beta
H_0}\}$. The Hamiltonian $H_0$ describes a system where the
original lattice has been reduced to a central site and to two
unconnected sublattices. Thus, in the $H_0$-representation, the
correlation functions connecting sites belonging to disconnected
sublattices can be decoupled. Within this scheme, the unknown
parameters can be written as functions of only two parameters $X_1
=\langle n^\alpha (i)\rangle _0 $ and $X_2 =\langle D^\alpha
(i)\rangle _0$, in terms of which one may find a solution of the
model. By exploiting the translational invariance along the chain,
one can impose $\langle n(i)\rangle =\langle n^\alpha (i)\rangle$
and $\langle D(i)\rangle =\langle D^\alpha (i)\rangle$, finding,
thus, two equations allowing one to determine $X_1$ and $X_2$ as
functions of $\mu$:
\begin{equation}
\label{eq43}
\begin{split}
 F_1 (X_1 ,X_2 ;\mu )&=0,
 \\
F_2 (X_1 ,X_2 ;\mu )&=0.
\end{split}
\end{equation}
The chemical potential $\mu $ can be determined by means of the
equation
\begin{equation}
\label{eq44} n=F_3 (X_1 ,X_2 ;\mu ).
\end{equation}
The explicit expressions of the functions $F_1$, $F_2$, and $F_3$
are given in the Appendix. Equations \eqref{eq43} and \eqref{eq44}
constitute a system of coupled equations allowing us to ascertain
the three parameters $\mu$, $X_1$ and $X_2$ in terms of the
external parameters of the model $n$, $h$, $U$, $V$, and $T$. Once
these quantities are known, all the properties of the model can be
computed.
\begin{figure}[t]
\centerline{\includegraphics[scale=0.65]{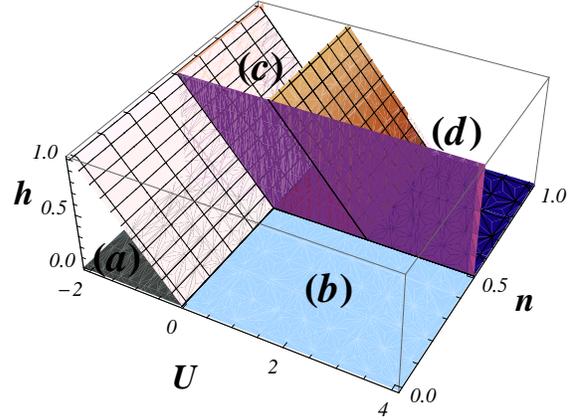}}
\caption{\label{fig1}The phase diagram in the space $(U,n,h)$ at
$T=0$ and $V=1$. The four different phases are characterized by
different distributions of the particles. Phase ($a$) is
characterized by zero magnetization and by only non-neighboring
doubly occupied sites. In the phase ($b$) the electrons singly
occupy non-neighboring sites with all the spins parallel to $h$.
In the phase ($c$) one finds both singly polarized and doubly
occupied non-neighboring sites, whereas in the phase ($d$)
neighboring sites can be occupied but no doublons are to be
found.}
\end{figure}

\section{The zero-temperature phase diagram}
\label{sec_III}

In this Section we derive the phase diagram of the AL-EHM in the
$(U,n,h)$ space. By numerically solving the set of equations
\eqref{eq43} and \eqref{eq44}, we study the $T=0$ behavior of
relevant physical quantities. This investigation  allows us to
envisage the distribution of the particles along the chain, for
different densities, as well as the magnetic properties of the
system. The results obtained are displayed in Fig. \ref{fig1}: one
recognizes the already known \cite{mancini2} four different phases
found at $h=0$ extending in the $h$ direction.

The phase structure is determined by the three competing terms of
the Hamiltonian: the repulsive intersite potential (disfavoring
the occupation of neighboring sites), the magnetic field (aligning
the spins along its direction, disfavoring thus double occupancy)
and the on-site potential, which can be either attractive or
repulsive. According to the values of these competing terms, one
may distinguish the different phases, characterized  by different
values of the double occupancy, the chemical potential, and of the
parameters defined in Eq. \eqref{eq35}.

\begin{figure}[t]
\centerline{\includegraphics[scale=0.29]{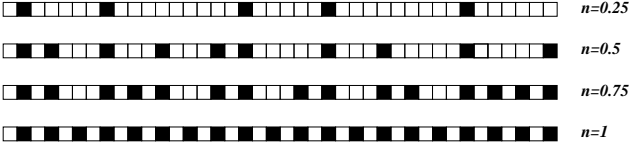}}
\caption{\label{fig2} Distribution of the particles along the
chain by varying the particle density at $T=0$, $V=1$, $U<0$ and
$h<\vert U \vert /2$. White and black squares denote empty and
doubly-occupied sites, respectively.}
\end{figure}

\begin{figure}[t]
\centerline{\includegraphics[scale=0.29]{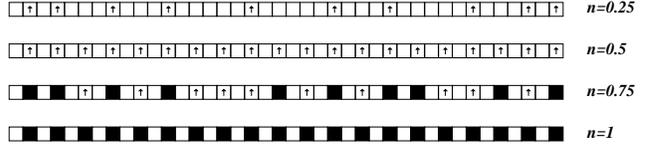}}
\caption{\label{fig3} Distribution of the particles  along the
chain by varying the particle density at $T=0$ and $V=1$. For $0<n
\le 0.5$, this distribution is observed both for $\{U<0$, $h>\vert
U \vert /2 \}$ and $\{ U>0$, $\forall h\}$. For $0.5<n \le 1$, it
corresponds to the regions $\{  U<0$, $\vert U \vert /2<h<V+\vert
U \vert /2 \}$ and $\{  0<U<2V$, $h< V-U/2 \}$. White and black
squares denote empty and doubly-occupied sites, respectively. The
squares with the arrow denote singly occupied sites with
field-aligned spins.}
\end{figure}

Phase ($a$) is observed in the region $0 \le n \le 1$ for
attractive on-site potential ($U<0$) and $h<\vert U \vert /2$; it
is just a continuation along the $h$ axis of the phase observed at
$h=0$ \cite{mancini2} and is characterized by a zero magnetization
$m=0$. The parameters $X_1$ and $X_2$ take the values
$X_1=2X_2=2n/(2-n)$. The chemical potential takes the value $U /2$
for $n<1$, whereas at $n=1$ (half filling) it jumps to the value
$\mu =2V+U/2$, as required by particle-hole symmetry. The
distribution of the particles in the phase ($a$) is shown in Fig.
\ref{fig2}, where we report just one possible configuration. The
attractive on-site potential favors the formation of doubly
occupied sites. At the same time, the nearest-neighbor repulsion
$V$ disfavors the occupation of neighboring sites. This
distribution of the electrons is confirmed by the expectation
values $D=n/2$ and $\lambda^{(k)} =0$. When $n<1$, there is no
ordered pattern in the distribution of the particles, whereas, for
$n=1$, one observes the well-known checkerboard distribution of
doubly-occupied sites \cite{mancini2,Bari71}.
The phase ($b$) is observed only for particle densities equal or
less than quarter filling ($n=0.5$), in the regions ($U>0$,
$\forall h$) and ($U<0$, $h>\vert U \vert/2$). In the latter
region, there is a competition between the attractive on-site
potential which favors the creation of doublons and the magnetic
field which favors the alignment of the electrons. When $h>\vert U
\vert/2$, it is energetically convenient to singly occupy
non-neighboring sites. Similarly, for $U>0$, the repulsion between
electrons on the same site and on neighboring sites ($V>0$), leads
to a scenario where the double occupancy, as well as the
short-range correlation functions $\lambda^{(k)}$, vanishes in the
limit $T \to 0$. One observes that the electrons tend to singly
occupy non-neighboring sites with all the spins parallel to $h$,
leading to a finite magnetization $m=n$. For particle densities
less than quarter filling, there is a cost in energy to add one
electron which is proportional to the intensity of the magnetic
field: $\mu=-h$. At $n=0.5$, the chemical potential jumps to the
value $\mu =V-h$ and a long-range order is established: one
observes a checkerboard distribution of singly occupied polarized
sites, as evidenced in Fig. \ref{fig3}. In this region one finds
$X_1=n/(1-n)$ and $X_2=0$.

\begin{figure}[ht]
\centerline{\includegraphics[scale=0.29]{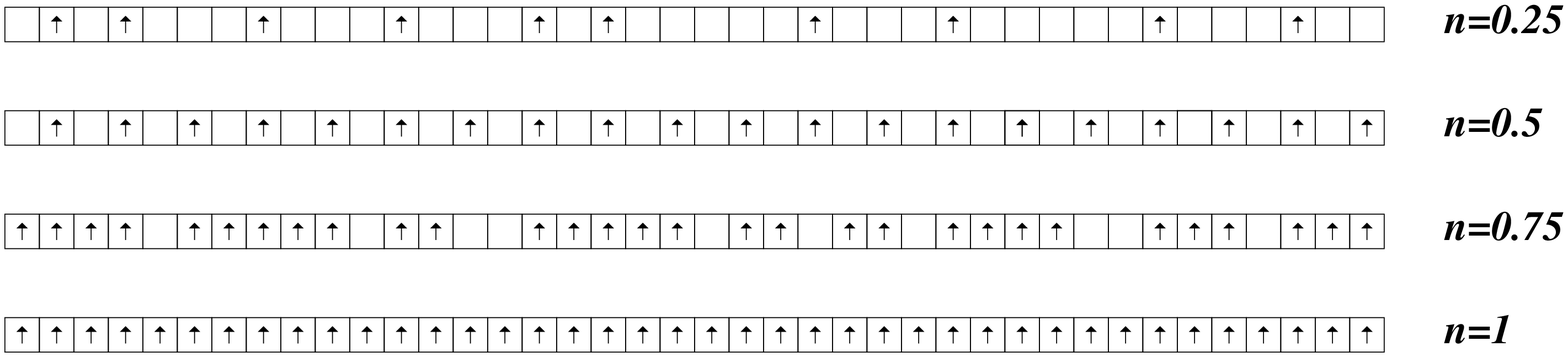}}
\caption{\label{fig4} Distribution of the particles  along the
chain by varying the particle density at $T=0$ and $V=1$. This
distribution is observed in two different parameters regions,
namely $\{ U<2V$, $h>V+ U/2 \}$ and $ \{U>2V, \forall h \}$. White
squares denote empty sites, whereas the squares with the arrow
denote singly occupied sites with field-aligned spins.}
\end{figure}

For particle densities greater than quarter filling, the phase
diagram is richer in the plane ($U,h$): one observes three phases
by varying $U$ ($h$) and keeping $h$ ($U$) fixed, as evidenced in
Fig. \ref{fig1}. Phase ($a$) has been discussed above. The phase
($c$) is observed for $0.5< n\le 1$ and in the regions ($0<U<2V$,
$h<V-U/2$)  and ($U<0$, $\vert U \vert/2<h<V+\vert U \vert/2$).
The chemical potential takes the value $U+h$ for $0.5< n<1$ and
$2V+U/2$ at $n=1$. In this region the intersite interaction
dominates: the minimization of the energy requires the extra
electrons above $n=0.5$ to occupy non-empty sites. Thus, in the
thermodynamic limit, there can be singly polarized and doubly
occupied sites but no neighboring sites: $\langle n(i)[n^\alpha
(i)]^k\rangle =0$. By increasing $n$, the number of
doubly-occupied sites increases linearly, $D=n-1/2$, while the
magnetization decreases as $m=1-n$. The distribution of the
particles in the region $(c)$ is drawn in Fig. \ref{fig3}, where
we represent just one possible configuration. The ground states of
the Hamiltonian are checkerboard configurations where empty sites
alternate with sites occupied by either two particles or one
particle with polarized spin. At half filling, one observes a
checkerboard distribution of only doubly occupied sites
\cite{Bari71}.
Also the phase ($d$) is observed for both attractive and repulsive
on-site interactions. In particular, when $U>2V$ the on-site
interaction dominates over the nearest-neighbor repulsion $V$. The
minimization of the energy requires the electrons not to be paired
and allows for the occupation of neighboring sites: $D=0$ and
$\lambda^{(1)}  =n-1/2$. The combined action of $U$ and $h$
predominates over $V$, leading to the same distribution in the
region ($0<U<2V$, $h>V-U/2$). For $U<0$, a strong magnetic field
($h>V+\vert U \vert/2$) dominates both the on-site and intersite
potentials, resulting on the absence of doublons and the
possibility to have neighboring sites occupied. In this phase, one
finds $X_1=1$ and $X_2=0$. For $n<1$ the energy necessary to add
one electron is $\mu =2V-h$. At $n=1$ the chemical potential jumps
to the value $\mu =2V+U/2$. Since in this state every occupied
site contains one and only one polarized electron, the ground
state is ferromagnetic and the magnetization is $m=n$, as
evidenced in Fig. \ref{fig4}.

\begin{figure}[t]
\centerline{\includegraphics[scale=0.88]{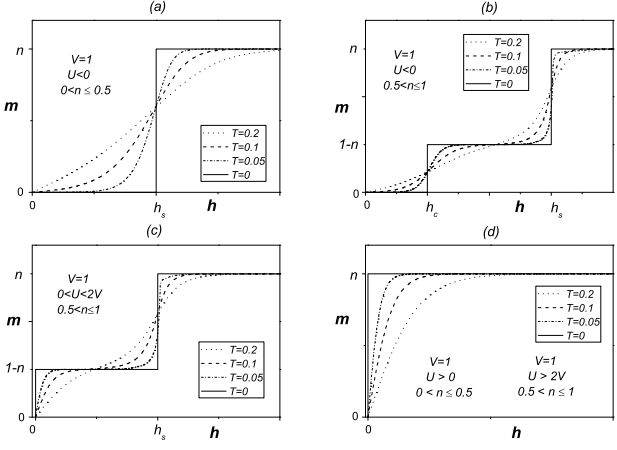}}
\caption{\label{fig5} The magnetization $m$ as a function of the
magnetic field $h$ at $V=1$ and decreasing temperatures.}
\end{figure}

These peculiar distributions of the electrons along the chain by
varying the magnetic field give rise to the formation of plateaus
in the magnetization curves. By increasing the magnetic field,
there are plateaus whose starting points depend on the particle
density, as well as on the on-site potential: one identifies two
critical values of the magnetic field. The nonzero magnetization
can either begin from $h=0$ or from a finite field. $h_c$ denotes
the starting point of a nonzero magnetization, whereas $h_s$
denotes the value of the magnetic field when it reaches
saturation. The results for the magnetization $m(h)$ are shown in
Fig. \ref{fig5}. The values of $h_c$ and $h_s$ can be inferred
from the previous analysis. Making  reference to Fig. \ref{fig5}
for the different regions of $n$ and $U$, one has: $h_c=h_s=\vert
U \vert /2$ (Fig. \ref{fig5}a), $h_c=\vert U \vert /2$ and
$h_s=V+\vert U \vert /2$ (Fig. \ref{fig5}b), $h_c=0$ and $h_s=V-U
/2$ (Fig. \ref{fig5}c), and $h_c=h_s=0$ (Fig. \ref{fig5}d).

The so-called metamagnetic behavior is clearly seen: at low
temperatures the magnetization begins to show a typical
"$S$-shape" which becomes more pronounced by further decreasing
the temperature. At $T=0$ one, two or three plateaus ($m=0$,
$m=1-n$ and $m=n$) are observed, according to the values of the
external parameters.
These results are similar to the ones obtained in a
one-dimensional spin-1 antiferromagnetic Ising chain with
single-ion anisotropy \cite{chen03,mancini08}.

\begin{figure}[th]
\centerline{\includegraphics[scale=0.88]{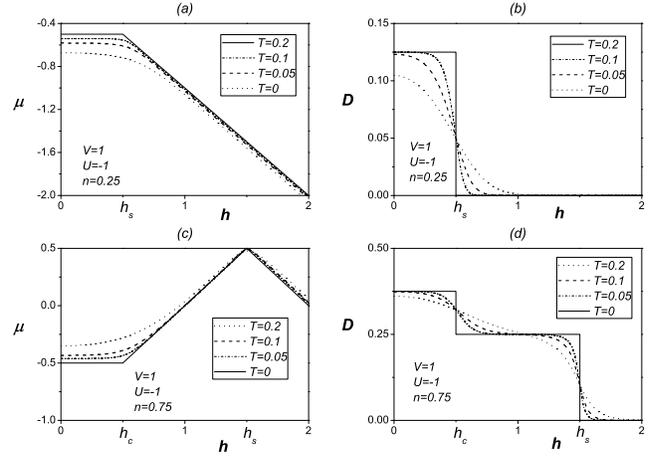}}
\caption{\label{fig6} The chemical potential and the double
occupancy at $V=1$, $U=-1$ and different values of the
temperature, for $n=0.25$ and $n=0.75$.}
\end{figure}

In Figs. \ref{fig6} we plot the double occupancy and the chemical
potential as functions of $h$ for $U=-1$, $n=0.25$, 0.75 and
different temperatures. As the temperature decreases, $D$ exhibits
sharp jumps, whereas $\mu$ shows discontinuity(ies). In
particular: (i) for all values of the particle density, $\mu$
takes the constant value $\mu=U/2$ in the region $U<0$ and
$h<\vert U \vert /2$, corresponding to $h_s$ ($h_c$) for $n=0.25$
(0.75); (ii) for $0<n<0.5$ and $h>h_s$, $\mu$ decreases with the
law $\mu=-h$. On the other hand, for $0.5<n<1$, $\mu$ first
increases with the law $\mu=U+h$ until $h_s=V+\vert U \vert /2$.
Further increasing $h$ one observes a decrease of the chemical
potential following the law $\mu=2V-h$. As it is evident from
Figs. \ref{fig6}, at $T=0$ and for repulsive on-site interactions,
the double occupancy shows a one or a two step-like behavior
according to the particle density. The steps occur at the values
of the magnetic field where PTs are observed.
In closing this Section, it is worthwhile to mention that at $T=0$
and $h=0$ all the phases (with the exception of the case $n=1$ and
$U<2V$) exhibit a macroscopic degeneracy growing exponentially
with the volume of the lattice. A nonzero magnetic field can lift
the degeneracy of the ground states at quarter and half filling,
even for repulsive on-site interactions.

\section{Finite temperature}
\label{sec_IV}

In this Section we shall investigate the finite temperature
properties of the AL-EHM. We study the behavior of the system as a
function of the parameters $n$, $T$, $U$ and $h$, and again we
take $V=1$ as the unit of energy and we set the Boltzmann's
constant $k_B =1$.

\subsection{Thermal properties}

\begin{figure}[b]
\centerline{\includegraphics[scale=0.88]{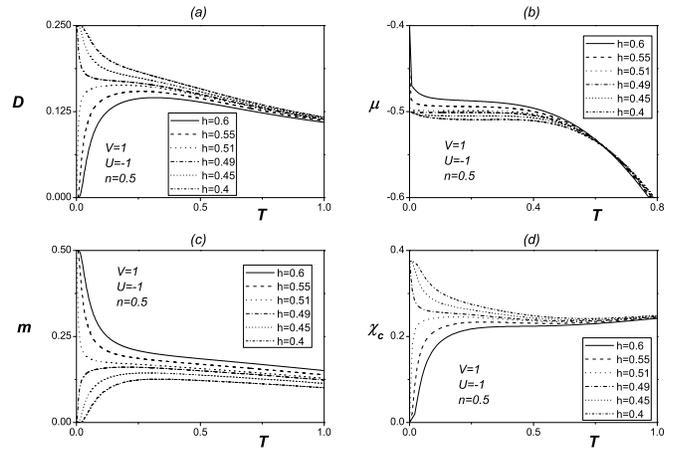}}
\caption{\label{fig7} (a) The double occupancy, (b) the chemical
potential, (c) the magnetization $m$, and (d) the charge
susceptibility as  functions of the temperature for $V=1$, $U=-1$
and $n=0.5$ in the neighborhood of $h_s=0.5$, where the transition
from phase ($a$) to phase ($b$) occurs.}
\end{figure}

In  Fig. \ref{fig7} we show the behavior of different quantities
as a function of $T$ for $n=0.5$, $U=-1$ and various values of $h$
in the vicinity of the critical point $h_s= \vert U \vert /2=0.5$.
These behaviors are not peculiar to the given values of the
parameters, but are always observed in the neighborhood of all
critical values of $h$ when the system goes from one phase to
another.
In the high temperature regime the double occupancy, the chemical
potential and the magnetization decrease by increasing $T$,
whereas for $T \to 0$ they tend to two different constants, as
shown in Figs. \ref{fig7}. At $h=h_s$ and $T=0$ there is a phase
transition: for the considered values of the external parameters,
one finds that the double occupancy jumps from zero to $n/2$, the
magnetization from zero to $n$ and the chemical potential from
$U/2$ to $V-h$. Also the charge susceptibility, defined as
$\chi_c=N~n^2+k_B~T$ $
\partial n/\partial \mu$, behaves differently at low
temperatures  according to the value of $h$. For $h<h_s$ the
charge susceptibility increases by decreasing $T$ and tends to
$\chi_c =n(1-n)(2-n)$ at $T=0$. For $h>h_s$ the charge
susceptibility decreases by decreasing $T$ and tends to $\chi_c
=n(1-n)(1-2n)$ at $T=0$. In the particular case shown in Fig.
\ref{fig7}d, $\chi_c$ vanishes for $T \to 0$ when $h>h_s$: the
system is in a charge ordered state \cite{mancini2}. The
non-vanishing of the charge susceptibility in the limit $T \to 0$
corresponds to a non-ordered ground state. For $h=h_s$ and $T=0$
there is a phase transition and the charge susceptibility exhibits
the discontinuity $\Delta \chi _c =n(1-n^2)$.
\begin{figure}[t]
\centerline{\includegraphics[scale=0.88]{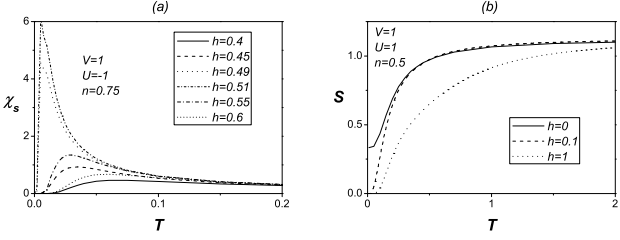}}
\caption{\label{fig8} (a) The spin susceptibility as a function of
the temperature for $V=1$, $U=-1$, $n=0.75$ in the neighborhood of
$h=\vert U \vert /2$. (b) The entropy $S$ as a function of the
temperature at $V=1$, $U=1$, $n=0.5$ and different values of $h$.}
\end{figure}

Another signature of the $T=0$ transition in the neighborhood of
the critical values of the magnetic field is provided by the
behavior of the spin susceptibility, defined as $\chi_s=\partial
m/\partial h$. As an example, in Fig. \ref{fig8}a, we plot
$\chi_s$ for $U=-1$, $n=0.75$ and different values of $h$ around
$h_c=\vert U \vert /2$. In the high temperature regime the spin
susceptibility decreases by increasing $T$. In the low temperature
regimes, the spin susceptibility exhibits a peak at a temperature
$T_1$, then decreases going to zero at $T=0$. By approaching the
value $h=h_c$ (both from below and above) the position $T_1$ of
the peak moves towards lower temperatures; at the same time the
height $h_1$ of the peak increases. At $h=h_c$ the spin
susceptibility diverges at $T=0$.


Recently, we have discuss how the entropy may be computed when the
ground state is degenerate \cite{mancini2}. In this case, the
zero-temperature entropy $S(0)$ is non-zero and, if the system is
not confined in one of the possible phases, is not even constant
but depends on the external parameters. The magnetic field can
remove the $h=0$ ground state degeneracy, responsible of a finite
zero-temperature entropy. This can be clearly seen in Fig.
\ref{fig8}b where we plot the entropy as a function of the
temperature at $U=1$ and $n=0.5$ for increasing values of the
magnetic field. As soon as the magnetic field is turned on, the
entropy vanishes in the limit $T \to 0$.

\begin{figure}[t]
\centerline{\includegraphics[scale=0.88]{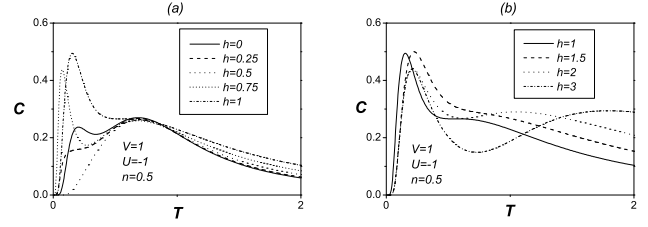}}
\caption{\label{fig9} The specific heat $C$ as a function of the
temperature for $V=1$, $U=-1$, $n=0.5$ and for different values of
$h$.}
\end{figure}

The study of the specific heat further enlightens the influence of
the magnetic field on the thermodynamic behavior of the system.
The specific heat is given by $C=dE/ dT$, where the internal
energy $E$ can be computed as the thermal average of the
Hamiltonian \eqref{eq5} and it is given by $E=UD+2V\lambda
^{(1)}-h m$. As an example of the characteristic behavior of the
specific heat by varying $h$, in Fig. \ref{fig9} we plot $C$ as a
function of the temperature at $V=1$, $U=-1$, and $n=0.5$, for
several values of the magnetic field. With the exception of the
critical value $h=h_s$, one observes an exponential activation of
$C$ with a pronounced peak, whose position depends on $h$.
Furthermore, the low temperature peak observed at $n<1$ for
pertinent values of the on-site potential when $h=0$, tends to
disappear by approaching $h_{s}$, both from above and below. The
position of the peak moves towards lower temperatures and vanishes
exactly at $h=h_{s}$. Of course, also the low temperature peak
with exponentially increasing height (observed at half-filling in
the limit $U \to 2V$ at $h=0$ \cite{mancini2}) survives only for
small values of $h$, but  disappears at $h=h_s$. On the other
hand, for higher values of the magnetic field one always observes
a double peak structure in the specific heat for all values of the
particle density.
In this region, the low temperature peak remains constant by
varying $h$, whereas the high temperature broad peak moves towards
higher temperatures by increasing $h$.

Interestingly, a change from a single to a double-peak structure
by varying the magnetic field has also been observed in
one-dimensional ferrimagnets with alternating spins
\cite{maisinger98}.

\subsection{Magnetic properties}

A useful representation of the phase diagram is obtained by
plotting the thermodynamic quantities as functions of the magnetic
field $h$. Indeed, this representation constitutes another way to
detect the zero-temperature transitions from thermodynamic data.
In fact, at low temperatures, all the thermodynamic quantities
that we have investigated present minima, maxima, or
discontinuities in the neighborhood of the critical values of $h$
at which zero-temperature PTs occur.

\begin{figure}[t]
 \centering
\includegraphics[scale=0.88]{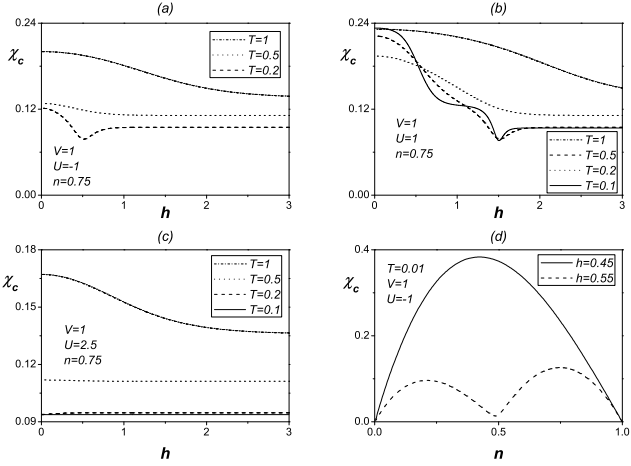}
 \caption{\label{fig10} The charge susceptibility  $\chi_c$ as a
function of the magnetic field at different temperatures for
$V=1$, $n=0.75$ and
 for (a) $U=-1$, (b)  $U=1$, and (c) $U=2.5$. (d) $\chi_c$ as a
function of the particle density for $V=1$, $U=-1$, $T=0.01$ and
different values of $h$.}
 \end{figure}

As a first example, in Figs. \ref{fig10}a-c we plot $\chi_c$ as a
function of the magnetic field, for $n=0.75$ and for different
values of the on-site potential ($U=-1$, 1, 2.5). In the high
temperature regime, the charge susceptibility is always a
decreasing function of $h$. In the limit $T \to 0$, one observes
two different behaviors, according to which phase the system is
in: $\chi_c$ tends to a constant when the system is in a
non-ordered phase, whereas it presents minima in correspondence of
the critical values of the magnetic field, as evidenced in
Fig.\ref{fig10}.
A similar behavior of the charge susceptibility is observed also
for other values of the particle density. In Fig. \ref{fig10}d we
plot $\chi_c$ as a function of the filling $n$ at $V=1$ and $U=-1$
and for two values of $h$ ($h=0.45$ and $h=0.55$, below and above
$h_{crit}=0.5$, respectively). To simplify notation, here we shall
use $h_{crit}$ to indicate either $h_c$ and $h_s$. One immediately
sees that, for $T \to 0$ and $h<h_{crit}$, the charge
susceptibility increases by increasing $n$ and has a maximum at
quarter filling. Further increasing $n$, $\chi _c$ decreases and
vanishes at half-filling, where, at fixed on-site potential, a
zero-temperature PT occurs by varying $n$, from a non-ordered to a
checkerboard distributions of doublons (see Fig. \ref{fig2}). When
$h>h_{crit}$, at low temperatures, $\chi_c$ has a double peak
structure with two maxima around $n=0.25$ and $n=0.75$ and two
minima around $n=0.5$ and $n=1$. This implies the presence of a
charge ordered state also at $n=0.5$, as evidenced by the charge
distribution shown in Fig. \ref{fig4}.

\begin{figure}[t]
\centerline{\includegraphics[scale=0.88]{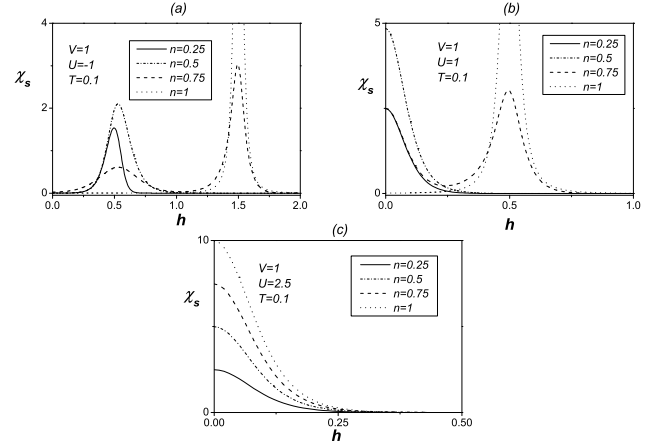}}
  \caption{\label{fig11} The spin susceptibility as a function of the magnetic field
  for $V=1$, $T=0.1$  and (a): $U=-1$, (b): $U=1$, (c): $U=2.5$.}
\end{figure}

In Figs. \ref{fig11}a-c we plot the spin susceptibility as a
function of the magnetic field at $T=0.1$ for three representative
values of $U$ ($U=-1$, $U=1$ and $U=2.1$, respectively) and for
different values of the filling ($n=0.25$, 0.5, 0.75, 1). The spin
susceptibility diverges in correspondence of the values $h_{crit}$
at which one moves from one magnetization plateau to the other.
For low values of the magnetic field and attractive on-site
interactions - corresponding to Fig. \ref{fig11}a - the spin
susceptibility tends to vanish at low temperatures for all values
of the filling: at $T=0$ all electrons are paired and no alignment
of the spin is possible. By increasing $h$, the magnetic
excitations break some of the doublons inducing a finite
magnetization: $\chi_s$ has a peak, then decreases, the system
having entered the successive magnetic plateau. If $0.5<n<1$ then
another peak is observed, corresponding to the second jump of the
magnetization when $h$ reaches the saturated value $h_s$. On the
other hand, for repulsive on-site interactions, a very small
magnetic field induces a finite magnetization (with the exception
of $n=1$ when $U<2V$): $\chi_s$ has a maximum at $h=0$ and
decreases by augmenting $h$, unless another transition line is
encountered, as it happens for $0.5<n \le 1$ and $0<U<2V$.

\begin{figure}[t]
\centerline{\includegraphics[scale=0.88]{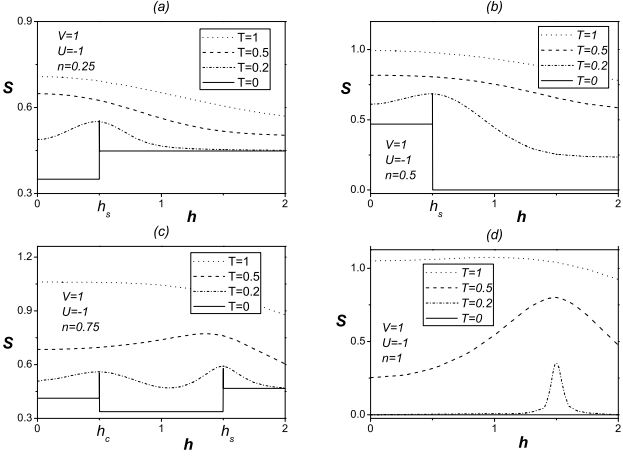}}
  \caption{\label{fig12} The entropy $S$ as a function of the magnetic field
  for $V=1$, $U=-1$, $T=1$, 0.5, 0.2  and (a): $n=0.25$, (b): $n=0.5$, (c): $n=0.75$, and (d): $n=1$.}
\end{figure}

In Figs. \ref{fig12} we plot the entropy as a function of the
magnetic field for relevant values of the particle density $n$.
The entropy presents maxima in the neighborhood of the values of
$h$ at which one observes zero-temperature PTs, and $S(0)$ has
discontinuities right at those values. On the other hand, for
sufficiently strong magnetic fields, the entropy becomes rather
insensitive to variations in $h$. When $n \le 0.5$, the entropy
presents a peak around $h_s$ which becomes more pronounced as the
temperature decreases. In the region $0.5<n<1$, at low
temperatures, one finds two peaks at $h_c$ and $h_s$,
respectively. Although at half filling  there is a
zero-temperature transition at $h_s$, one always finds $S(0)=0$
since the relative ground states are non-degenerate.

\section{Concluding remarks}
\label{sec_V}

We have evidenced how the use of the Green's function and
equations of motion formalism leads to the exact solution of the
one-dimensional AL-EHM limit in the presence of an external
magnetic field. We provided a systematic analysis of the model for
nearest-neighbor repulsion $V$ by considering relevant
thermodynamic quantities in the whole space of the parameters $n$,
$T$, $h$ and $U$ (having chosen $V=1$ as the unity of energy).
This study has shown that, at zero-temperature, the model exhibits
phase transitions for specific values of $n$, $h$ and $U$. In
particular, we have identified four phases in the ($U,n,h$) space
and PTs are observed at the borders of these phases. Various types
of long-range charge ordered states have been observed: (i) at
half-filling for $U<2V$ and $h<V-U/2$, a checkerboard distribution
of doubly-occupied sites; (ii) at quarter-filling for $U>0$, as
well as for $U<0$ and $h> \vert U \vert /2$, a checkerboard
distribution of polarized singly occupied sites; (iii) for $0.5\le
n \le 1$ an ordered state with alternating empty and occupied
sites in the regions ($U<0$, $\vert U \vert /2<h <V+\vert U \vert
/2$) and ($0<U<2V$, $h>V-U/2$).

We derived the phase diagram in the space $(U,n,h)$ at $T=0$ by
computing several quantities whose behaviors is also useful to
characterize the distribution of the electrons on the sites of the
chain. When plotted as functions of $h$, the chemical potential,
the double occupancy as well as the magnetization, show
discontinuities where PTs occur. We identified the values of the
critical fields $h_c$ and $h_s$, defining, according to the value
of the particle density, the beginning point of nonzero
magnetization and the saturated magnetization field, respectively.
Furthermore, the presence of the magnetic field may dramatically
modify the behavior of several thermodynamic quantities. For
instance, the charge susceptibility tends to different values in
the limit $T \to 0$, depending on how strong is the magnetic
field. At quarter and half filling, a strong magnetic field can
also remove the macroscopic degeneracy of the ground states: as a
result, the zero-temperature entropy is zero.

\appendix

\section{Computational details}
\label{app_A}

Firstly, we shall provide some more details on the matrices used
in the computational framework adopted in Sec. \ref{sec_II}. To
begin with, we note that, due to the spin conservation, the
normalization matrix $I^{(a)}=\langle \{\psi ^{(a)}(i,t),{\psi
^{(a)}}^\dag (i,t)\}\rangle$ is block diagonal:
\begin{equation}
I^{(a)}=\left( {{\begin{array}{*{20}c}
 {I^{\uparrow \uparrow ,(a)}}  & 0  \\
 0  & {I^{\downarrow \downarrow ,(a)}}  \\
\end{array} }} \right).
\end{equation}
The time translational invariance requires the $m$-matrix
$m=\langle \{i\frac{\partial }{\partial t}\psi (i,t),\psi ^\dag
(i,t)\}\rangle$ to be symmetric. This requirement leads to find
that several elements of the normalization matrix are not
independent, and one needs to compute only the matrix elements
 $\{I_{1,p}\}$, and $\{I_{6,p+6}\}$, where $p=1,\ldots,5$.
By recalling the basic commutators
\begin{equation}
\begin{split}
\{ \xi_{\sigma} (i),\xi_{\sigma}^\dag
(i)\}&=[1-n_{-\sigma } (i)], \\
 \{ \eta_{\sigma} (i),\eta_{\sigma}^\dag
(i)\}&=n_{-\sigma } (i),
 \end{split}
\end{equation}
one easily obtains Eq. \eqref{eq36}.


The matrices $\Omega ^{(\xi )}$ and $\Omega ^{(\eta )}$ have the
expression
\begin{equation}
\Omega ^{(\xi )}=\Omega ^{(\eta )}=
\begin{scriptstyle}
\begin{pmatrix}
\scriptstyle   1  & \scriptstyle  0  & \scriptstyle  {2^4}  &
\scriptstyle  0  & \scriptstyle  1  & \scriptstyle  0  &
\scriptstyle
{\big(\frac{2}{3}\big)^4}  & \scriptstyle  0  & \scriptstyle  {\big(\frac{1}{2}\big)^4}  & \scriptstyle  0  \\
 \scriptstyle  0  & \scriptstyle  0  & \scriptstyle  {2^3}  & \scriptstyle  0  & \scriptstyle  1  & \scriptstyle  0  & \scriptstyle
{\big(\frac{2}{3}\big)^3}  & \scriptstyle  0  & \scriptstyle  {\big(\frac{1}{2}\big)^3}  & \scriptstyle  0  \\
 \scriptstyle  0  & \scriptstyle  0  & \scriptstyle  {2^2}  & \scriptstyle  0  & \scriptstyle  1  & \scriptstyle  0  & \scriptstyle
{\big(\frac{2}{3}\big)^2}  & \scriptstyle  0  & \scriptstyle  {\big(\frac{1}{2}\big)^2}  & \scriptstyle  0  \\
 \scriptstyle  0  & \scriptstyle  0  & \scriptstyle  {2^1}  & \scriptstyle  0  & \scriptstyle  1  & \scriptstyle  0  & \scriptstyle
{\big(\frac{2}{3}\big)^1}  & \scriptstyle  0  & \scriptstyle  {\big(\frac{1}{2}\big)^1}  & \scriptstyle  0  \\
 \scriptstyle  0  & \scriptstyle  0  & \scriptstyle  {2^0}  & \scriptstyle  0  & \scriptstyle  1  & \scriptstyle  0  & \scriptstyle
{\big(\frac{2}{3}\big)^0}  & \scriptstyle  0  & \scriptstyle  {\big(\frac{1}{2}\big)^0}  & \scriptstyle  0  \\
\scriptstyle   0  & \scriptstyle  1  & \scriptstyle  0  &
\scriptstyle  {2^4}  & \scriptstyle  0  & \scriptstyle  1  &
\scriptstyle  0
 & \scriptstyle  {\big(\frac{2}{3}\big)^4}  & \scriptstyle  0  & \scriptstyle  {\big(\frac{1}{2}\big)^4}  \\
 \scriptstyle  0  & \scriptstyle  0  & \scriptstyle  0  & \scriptstyle  {2^3}  & \scriptstyle  0  & \scriptstyle  1  & \scriptstyle  0
 & \scriptstyle  {\big(\frac{2}{3}\big)^3}  & \scriptstyle  0  & \scriptstyle  {\big(\frac{1}{2}\big)^3}  \\
 \scriptstyle  0  & \scriptstyle  0  & \scriptstyle  0  & \scriptstyle  {2^2}  & \scriptstyle  0  & \scriptstyle  1  & \scriptstyle  0
 & \scriptstyle  {\big(\frac{2}{3}\big)^2}  & \scriptstyle  0  & \scriptstyle  {\big(\frac{1}{2}\big)^2}  \\
 \scriptstyle  0  & \scriptstyle  0  & \scriptstyle  0  & \scriptstyle  {2^1}  & \scriptstyle  0  & \scriptstyle  1  & \scriptstyle  0
 & \scriptstyle  {\big(\frac{2}{3}\big)^1}  & \scriptstyle  0  & \scriptstyle  {\big(\frac{1}{2}\big)^1}  \\
 \scriptstyle  0  & \scriptstyle  0  & \scriptstyle  0  & \scriptstyle  {2^0}  & \scriptstyle  0  & \scriptstyle  1  & \scriptstyle  0
 & \scriptstyle  {\big(\frac{2}{3}\big)^0}  & \scriptstyle  0  & \scriptstyle
 {\big(\frac{1}{2}\big)^0}
\end{pmatrix}.
 \end{scriptstyle}
\end{equation}


As it has been shown in Sec. \ref{sec_II}, the Green's and the
correlation functions depend on 13 parameters: $\mu$, $m$, $\kappa
^{(2)}$, $\kappa ^{(3)}$, $\kappa ^{(4)}$, $\lambda ^{(1)}$,
$\lambda ^{(2)}$, $\lambda ^{(3)}$, $\lambda ^{(4)}$, $\pi
^{(1)}$, $\pi ^{(2)}$, $\pi ^{(3)}$, $\pi ^{(4)}$.
By recalling the expression of  $H_I$ [see Eq. \eqref{eq39}], and
by using the recurrence rule \eqref{eq7},  it is easy to show that
\begin{equation}
\label{eqd4}
 e^{-\beta H_I }=1+\sum_{m=1}^4  \left[n(i)f_m +D(i) g_m \right] [n^\alpha (i)]^m ,
\end{equation}
where
\begin{equation}
\begin{split}
 f_m &=\sum_{n=1}^\infty (-1)^n\frac{1}{n!}A_m^{(n)} (2\beta V)^n, \\
 g_m &=\sum_{n=2}^\infty (-1)^n\frac{1}{n!}a_n A_m^{(n)} (2\beta
 V)^n.
\\
\end{split}
\end{equation}
The coefficients $f_m$ and $g_m$ are polynomials of $K=e^{-\beta
V}$ given by
\begin{equation}
\begin{split}
 f_1 &=-\frac{K^4}{2}+\frac{8K^3}{3}-6K^2+8K-\frac{25}{6} ,\\
 f_2 &=\frac{1}{6} (K-1)^2 \left(11 K^2-34 K+ 35 \right),
 \\
 f_3 &= -\frac{2}{3} (K-1)^3 (3 K-5),
  \\
 f_4 &=\frac{2}{3}(K-1)^4 ,
 \end{split}
 \end{equation}
 and
\begin{equation}
\begin{split}
 g_1
&=  -\frac{1}{6} (K-1)^5 \left(3 K^3+15 K^2+29 K+25\right),
\\
  g_2&= \frac{1}{6}
(K-1)^2
 \left( 11 K^6 + 22 K^5 - 23 K^4 - 68 K^3 \right.
 \\
 &-  \left. 21 K^2+ 138 K -35  \right),
\\
 g_3 &=  -\frac{2}{3} (K-1)^3 \left(3 K^5 +9 K^4 +4 K^3 \right.
 \\
 &-  \left.  12 K^2 -21 K+5\right),
 \\
 g_4&=\frac{2}{3} (K-1)^4 \left(K^4 +4 K^3 +6 K^2 +4 K
 -1\right).
 \end{split}
 \end{equation}
Now, by taking the expectation value of Eq. \eqref{eqd4} with
respect to $H_0$, and exploiting to the properties of the $H_0
$-representation, one finds
\begin{equation}
\label{eqd8}
\begin{split}
\Upsilon_0 &=\langle e^{-\beta H_I }\rangle _0 =1+\langle
n(i)\rangle _0
\sum_{m=1}^4 f_m \langle [n^\alpha (i)]^m\rangle _0 \\
&+\langle D(i)\rangle _0 \sum_{m=1}^4 g_m \langle [n^\alpha
(i)]^m\rangle _0 .
\end{split}
 \end{equation}
In order to compute the quantities $\langle n(i)\rangle _0$ and
$\langle D(i)\rangle _0 $, one may use the equations of motion
\begin{equation}
\begin{split}
[\xi _\sigma (i),H_0 ]&=-(\mu +\sigma h)\xi _\sigma (i) ,
\\
[\eta _\sigma (i),H_0 ]&=-(\mu +\sigma h-U)\eta _\sigma (i),
\end{split}
\end{equation}
leading, for a homogeneous phase, to the following expressions for
the CFs:
\begin{equation}
\label{eqd13}
\begin{split}
C_\sigma ^{(\xi ,0)} &=\langle \xi _\sigma (i)\xi _\sigma ^\dag
(i)\rangle _0
=\frac{1-\langle n_{-\sigma } (i)\rangle _0 }{1+e^{\beta (\mu +\sigma h)}} ,\\
C_\sigma ^{(\eta ,0)} &=\langle \eta _\sigma (i)\eta _\sigma ^\dag
(i)\rangle _0 =\frac{\langle n_{-\sigma } (i)\rangle _0
}{1+e^{\beta (\mu +\sigma h-U)}}.
\end{split}
\end{equation}
Recalling that $\xi _\sigma \xi _\sigma ^\dag +\eta _\sigma \eta
_\sigma ^\dag =1-n_\sigma$ and  $\eta _\sigma \eta _\sigma ^\dag
=n_{-\sigma } -n_\uparrow n_\downarrow$, from Eq. \eqref{eqd13}
one obtains
\begin{equation}\begin{split}
 B_1^\uparrow &=\langle n_\uparrow (i)\rangle _0 =\frac{e^{\beta (\mu +2h)}+e^{\beta (2\mu
+h-U)}}{e^{\beta h}+e^{\beta \mu }+e^{\beta (\mu +2h)}+e^{\beta
(2\mu +h-U)}}, \\
 B_1^\downarrow &=\langle n_\downarrow (i)\rangle _0 =\frac{e^{\beta \mu }+e^{\beta (2\mu
+h-U)}}{e^{\beta h}+e^{\beta \mu }+e^{\beta (\mu +2h)}+e^{\beta
(2\mu +h-U)}} ,\\
 B_1 &=B_1^\uparrow +B_1^\downarrow =\frac{e^{\beta \mu
}(1+e^{2\beta h}+2e^{\beta (\mu +h-U)})}{e^{\beta h}+e^{\beta \mu
}+e^{\beta (\mu +2h)}+e^{\beta (2\mu +h-U)}} ,\\
 B_2 &=\langle D(i)\rangle _0 =\frac{e^{\beta (2\mu +h-U)}}{e^{\beta h}+e^{\beta \mu
}+e^{\beta (\mu +2h)}+e^{\beta (2\mu +h-U)}}, \\
 B_3 &=B_1^\uparrow -B_1^\downarrow =\frac{e^{\beta \mu }(e^{2\beta
h}-1)}{e^{\beta h}+e^{\beta \mu }+e^{\beta (\mu +2h)}+e^{\beta
(2\mu +h-U)}}.
 \end{split}
 \end{equation}
Because of the properties of the $H_0$-representation, by using
Eq. \eqref{eqd4} and by means of the relations
\begin{equation}
\begin{split}
 n_\sigma (i)n(i)=n_\sigma (i)+D(i), \\
 n_\sigma (i)D(i)=D(i), \\
 \end{split}
\quad
\begin{split}
 n_3 (i)n(i)=n_3 (i), \\
 n_3 (i)D(i)=0 ,
 \end{split}
 \end{equation}
one finds
\begin{equation}
\label{eqd18}
\Upsilon_0  =1+\sum_{m=1}^4 (B_1 f_m +B_2 g_m
)\langle [n^\alpha (i)]^m\rangle _0,
\end{equation}
and
\begin{equation}
\begin{split}
 n&=\frac{1}{\Upsilon_0 }\{B_1 +\sum_{m=1}^4 [(B_1 +2B_2 )f_m
+2B_2 g_m ]\langle [n^\alpha (i)]^m\rangle _0 \}, \\
 m&=\frac{B_3 }{\Upsilon_0 }\{1+\sum_{m=1}^4 f_m \langle [n^\alpha
(i)]^m\rangle _0 \} ,\\
 D&=\frac{B_2 }{\Upsilon_0 }\{1+\sum_{m=1}^4 (2f_m +g_m
)\langle [n^\alpha (i)]^m\rangle _0 \}.
 \end{split}
 \end{equation}
Upon defining the two parameters
\begin{equation}
\begin{split}
 X_1 &=\langle {n^\alpha (i)} \rangle_0 ,\\
 X_2 &=\langle {D^\alpha (i)} \rangle_0  ,
 \end{split} \end{equation}
and by exploiting the translational invariance along the chain,
$\langle n(i)\rangle =\langle n^\alpha (i)\rangle$ and $
 \langle D(i)\rangle =\langle D^\alpha (i)\rangle$, one  obtains two equations
allowing one to determine $X_1$ and $X_2$ as functions of $\mu$,
namely:
\begin{equation}
\label{eqd43}
\begin{split}
F_1 &=X_1 -B_1 +\sum_{m=1}^4 (B_1 f_m +B_2 g_m )\langle [n^\alpha
(i)]^{m+1}\rangle _0 \\
&-\sum_{m=1}^4 [(B_1 +2B_2 )f_m +2B_2 g_m ]\langle [n^\alpha
(i)]^m\rangle _0 =0, \\
 F_2 &=X_2 -B_2 +\sum_{m=1}^4 (B_1 f_m +B_2 g_m )\langle D^\alpha
(i)[n^\alpha (i)]^m\rangle _0
\\
&-B_2 \sum_{m=1}^4 (2f_m +g_m )\langle [n^\alpha (i)]^m\rangle _0
=0.
\end{split}
\end{equation}
Upon defining $\Xi=(1-X_1 +X_2 )+(X_1 -2X_2 )(1+aX_1 +a^2X_2 )+X_2
(1+dX_1 +d^2X_2 )$, the chemical potential $\mu $ can be
determined by means of the equation
\begin{equation}
\begin{split}
\label{eqd44} F_3 &=\frac{1}{\Xi}(X_1 -2X_2 )(1+aX_1 +a^2X_2 )\\
&+2X_2 (1+dX_1 +d^2X_2)=n.
\end{split}
\end{equation}
After some lengthy but straightforward calculations, one finds
that Eqs. \eqref{eqd43} can be rewritten as
\begin{equation}
\label{eqd27}
\begin{split}
 X_1 &=2e^{\beta \mu }\cosh(\beta h)(1-X_1 -dX_2 )(1+aX_1 +a^2X_2 ) \\
 &+e^{\beta (2\mu -U)}[2+(d-1)X_1 -2dX_2 ](1+dX_1 +d^2X_2 ) ,\\
 X_2 &=e^{\beta (2\mu -U)}[1+dX_1 -(2d+1)X_2 ](1+dX_1 +d^2X_2 ) \\
 &-2e^{\beta \mu }\cosh(\beta h)(1+d)X_2 (1+aX_1 +a^2X_2 ).
 \end{split}
 \end{equation}
As a consequence, one also gets:
\begin{equation}
\label{eqd39}
\begin{split}
D&=\frac{1}{\Xi} X_2 (1+dX_1 +d^2X_2 ),
\\
m&=\frac{1}{\Xi} \tanh (\beta h)(X_1 -2X_2 )(1+aX_1 +a^2X_2)
 \\
\\
&=\tanh (\beta h)(n-2D) ,\\
 \lambda ^{(1)}&=\frac{1}{2\Xi} K[(X_1
-2X_2 )(X_1 +2aX_2 )
\\
&+2KX_2 (X_1 +2dX_2 )] ,
\end{split}
 \end{equation}
where
\begin{equation}
\begin{split}
 a&=e^{-\beta V}-1=K-1 ,\\
 d&=e^{-2\beta V}-1=K^2-1.
 \end{split}
 \end{equation}


We conclude this Appendix by reporting the analytical solution
obtainable at half filling. Particle hole symmetry requires that
at half filling the chemical potential must take the value $\mu
=U/2+2V$. For this value of $\mu$, Eqs. \eqref{eqd27} become
\begin{equation}
 \label{eqd45}
\begin{split}
X_1 &=
 2GK^{-2}(1-X_1 -dX_2 )(1+aX_1 +a^2X_2 ) \\
 &+K^{-4}[2+(d-1)X_1 -2dX_2 ](1+dX_1 +d^2X_2 ),\\
 X_2 &=K^{-4}[1+dX_1 -(2d+1)X_2 ](1+dX_1 +d^2X_2 ) \\
 &-2GK^{-2}(1+d)X_2 (1+aX_1 +a^2X_2 ) ,
 \end{split}
  \end{equation}
where $G=e^{\beta U/2}\cosh(\beta h)$. The solution of the above
equations is:
\begin{equation}
\label{eqd50}
\begin{split}
 X_1 &=1-\frac{(K+1)[1+2GK+K^2- Q]}{4GK(K-1)},
\\
 X_2 &=\frac{1+2GK+K^2-Q }{4GK(K-1)^2},
 \end{split}
  \end{equation}
where $Q=\sqrt {(1+2GK+K^2)^2-8GK(K-1)^2}$. Upon, inserting Eq.
\eqref{eqd50} in Eqs. \eqref{eqd44} and \eqref{eqd39} one finds
\begin{equation}
\begin{split}
n&=1 ,\\
D&=\frac{1}{2[1+G(1-a^2 X_2)^2]}, \\
 \lambda^{(1)}&=\frac{1-2a^2 X_2 + (1-4K+2K^2+K^4)X_2^2}{2\left[1-2a^2 X_2+a^2(1+K^2)X_2^2\right]}.
 \end{split}
  \end{equation}


\begin{thebibliography}{}




\bibitem{Mancini05_06} F. Mancini, Europhys. Lett. \textbf{70}, 484 (2005);
 Condens. Matter Phys. \textbf{9},  393 (2006).


\bibitem{mancini2} F. Mancini, F. P. Mancini, Phys. Rev. E  {\bf 77}, 061120 (2008).

\bibitem{manciniavella} F. Mancini, A. Avella, Adv. Phys. {\bf 53}, 537 (2004).

\bibitem{imambekov04} A. Imambekov, M. Lukin, E. Demler, Phys. Rev. Lett. {\bf 93}, 120405 (2004).

\bibitem{haldane} F. D. M. Haldane, Phys. Lett. A {\bf 93}, 464 (1993); Phys. Rev. Lett.
{\bf 50}, 1153 (1983).

\bibitem{mancini05b}  F. Mancini, Eur. Phys. J. B {\bf 47} 527, (2005).

\bibitem{Bari71} R. A. Bari, Phys. Rev. B \textbf{3}, 2662 (1971).

\bibitem{chen03} X. Y. Chen, Q. Jiang, W. Z. Shen, C. G. Zhong, Jour. Magn. Magn.
Mat. {\bf 262}, 258 (2003).

\bibitem{mancini08} F. Mancini, F. P. Mancini, Condens. Matter Phys. {\bf 11}, 543 (2008).

\bibitem{maisinger98} K. Maisinger, U. Schollw\"ock, S. Brehmer,
H. J. Mikeska, S. Yamamoto, Phys. Rev. B {\bf 58},  R5908 (1998).


\end{thebibliography}
\end{document}